# Thermal AGN Signatures in Blazars

**Eric S. Perlman[1], Brett Addison**
*Florida Institute of Technology*
*Melbourne, FL, USA*
*E-mail:* `eperlman@fit.edu`

**Markos Georganopoulos, Brian Wingert, Philip Graff**
*University of Maryland, Baltimore County*
*Baltimore, MD, USA*
*E-mail:* `georgano@umbc.edu`

Long ignored in blazars because of the dominance of the beamed radiation from the jet, the topic of thermal emissions in these objects is just beginning to be explored. While this emission is weak in most blazars compared to the dominant nonthermal jet components, there is a growing body of evidence that suggests that thermal emission is observable even in the most highly beamed objects. The emitting regions, which can include the accretion disk as well as the torus, are key parts of the central engine which also powers the jets. They also may be of critical importance in helping us decide between unified scheme models.

We will review the observational evidence for thermal emissions in blazars, with an emphasis on recent work, and the spectral and variability characteristics that have been observed. The majority of the evidence for thermal emission in blazars (now observed in several objects) has come as a result of multiwavelength campaigns, where the object showed a clear "bump" in the optical-UV in a faint state. However, evidence exists from other avenues as well, including both purely spectral and variability based arguments as well as statistical analyses of large samples of objects. We will also discuss the impact of thermal emission on an object's overall SED, including its Comptonized signatures. Finally, we will assess the current standing of unified scheme models as respects thermal signatures and the prospects for detecting thermal emission with new telescopes and missions, and further utilizing it as a probe of the central engine of blazars.

---

[1] Speaker

# 1. Introduction

Blazars, like all AGN, are complex animals. The central engine is believed to have a complex structure (Figure 1) that features an accretion disk circling a central, supermassive black hole, broad and narrow emission line regions, an obscuring structure sometimes referred to as the torus (although whether its geometry actually merits that name is a subject of recent, active debate not germane to this article). Each of these regions radiates in different bands, and is known to have different signatures. In blazars, the focus has historically been on various characteristics of the jet, including multiwaveband variability, jet morphology, polarization, etc. This is because of the preferred orientation of the jet represented in blazars, where as a result of relativistic beaming, the synchrotron and inverse-Compton emission from the jet can be amplified by large factors [1]. Precisely because of their jet-on orientation, the thermal and line emission regions that dominate the properties of other AGN classes, are less prominent in blazars. However, just because those characteristics are less prominent in blazars does not mean they are absent. In fact, for unified schemes to be successful, we must not only be able to duplicate the range of jet properties seen in blazars and tie them to the jet characteristics seen in the parent population(s) of radio galaxies, we must also be able to match the characteristics of emission sites outside of the jet, the radiation from which would emerge isotropically and would not be affected by Doppler boosting.

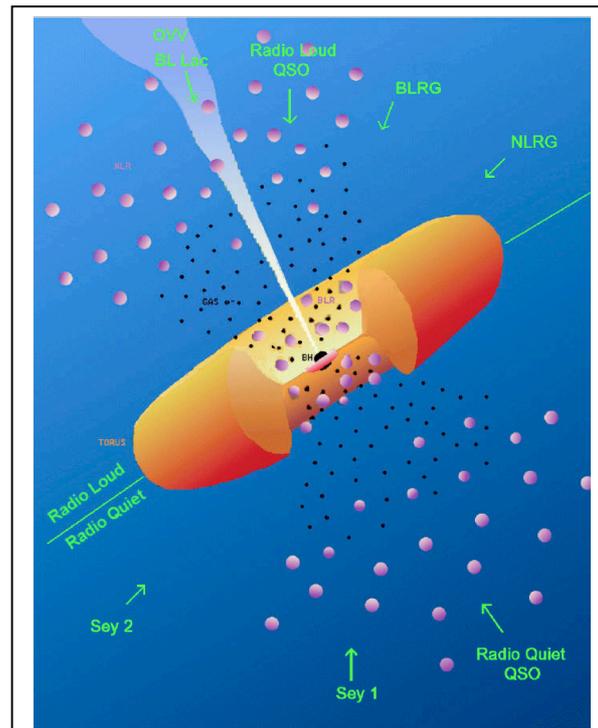

Figure 1. The Unified Model for active galactic nuclei. Note the multiple emission regions and the particular view afforded by blazars.

Thermal emission can emerge from different regions of the central engine of AGNs. Thermal continuum emission in the ultraviolet and optical is believed to originate within the accretion disk, while the thermal emission seen in the infrared is believed to originate in the dusty, molecular torus, representing emission from the disk or BLR that has been reprocessed in this optically thick region into thermal radiation (see e.g., [2] for a review). Moreover, the broad and narrow line emission which are the calling cards of other AGN classes are also themselves linked to the disk's thermal emission, as under unified schemes the jet only illuminates a small solid angle ($\sim 1/\Gamma^2$, where $\Gamma$ is the bulk Lorentz factor of the jet), and also does not represent the dominant source of ionizing UV and soft X-ray emission, when all emissions in those bands are integrated over $4\pi$ steradians. The anti-correlation between the core to extended radio flux ratio *R* (often used as a proxy for measuring the degree of beaming [3,4,5,6,7,8,9,10]) and broad-line equivalent width [11,12] is often taken as evidence of the lack of a link between the jet emission and that from the BLR. Thus the thermal emission regions are critical parts of the AGN structure under unified schemes, and furthermore they are the main power source for the emission line regions that characterize other AGN classes.

In blazars, the thermal continuum emission is much more difficult to locate than in other AGN classes, because of the orientation and the resultant relativistic beaming of the jet. Precisely how different the observed ratio between jet and thermal emission will be in blazars as compared to other AGN classes can be seen by imagining a putative object, in which the thermal emission from the disk comprises ~90% of the continuum that would be observed in the UV from an off-axis angle (typical of steep-spectrum radio quasars, [1,2]). If one then inserts into this putative object a jet with a rather modest Lorentz factor $\Gamma=5$ and views it at an angle where the Doppler boosting factor $\delta=\Gamma$, however, suddenly the apparent picture changes, with the nonthermal UV emission from the jet now dominating the observed continuum emission by more than a factor 60. Thus even in strong-lined blazars (i.e., FSRQs), thermal characteristics will be difficult to locate and characterize, and very often they may not be seen. In weaker-lined objects (i.e., BL Lacs), where the reduced line emission likely also indicates less prominent thermal emission from the disk and possibly also a less optically thick or absent torus, this emission will be even fainter and harder to locate, let alone characterize.

Despite the inherent difficulty in observing thermal emission in blazars, the importance of this thermal emission to unified schemes, makes it important to characterize thermal emission. The purpose of this paper is therefore to describe what evidence for thermal emission is seen in blazars, and more broadly describe how we should distinguish thermal continuum emission from nonthermal in blazars, either spectrally or via variability characteristics. This latter part of the paper is of key importance, because we must realize that the thermal emission from both the disk and torus does not typically represent a single temperature, so it is also a broadband component whose properties can, over multiple decades in frequency, approximate a power-law. However, as shown by the composite spectra of quasars that have been compiled in recent years [13,14], we can use the fact that the disk and torus emission should comprise relatively distinct 'bumps' due to radically different temperature and density characteristics, to attempt to isolate them. This also allows us to use some of the properties found in unbeamed quasars to tell us what we should expect to see as far as the variability characteristics of thermal emission.

We will undertake first to describe what one might call an 'ideal' situation in Section 2, and thus attempt to outline what characteristics we should be able to take as evidence of, or tracers of, thermal emission. In Section 3, we will describe the current state of observational evidence, beginning with a theoretical approach and then continuing to papers of the last few years. In assessing the more recent findings, we will also attempt to place a 'confidence' upon their detection, or lack thereof, of thermal emission. In Section 4, we will take this information and assess the impact that thermal emission would have on the higher-energy emissions from jets, e.g., by Comptonization. We will also assess unified scheme models in this section. We will close in Section 5 with a discussion of the prospects for further discovery afforded by future missions and telescopes.

## 2. What Should We Expect to See?

It is useful in this discussion to identify straight off the characteristics we should target in our search for thermal emission in blazars. This will aid us in separating characteristics that are *bona fide* linked to the thermal emission regions, from those which are not or may not be. We will start this description below by first describing what one might call the 'ideal' situation, and then move on to describe various complications caused by the orientation we observe in blazars.

By its very nature, thermal emission covers a much more narrow band than radiation generated by the synchrotron or inverse-Compton mechanisms. Thus one of the calling cards of thermal emission would be a 'bump' in the spectral energy distribution (SED) in a specific waveband, e.g., in the UV for accretion disk emission. The second thing we would expect ideally is that the variability characteristics of thermal emission should be entirely different from those seen in the nonthermal radiation. Not only should they have no link to those of the thermal emission, but the timescales should be entirely different. This can be seen by examining the variability characteristics of quasars that are not dominated by nonthermal emission (e.g., [15]), where what we see in the IR through UV is largely frequency-independent variability on timescales of weeks to months, with amplitudes of a magnitude or less, considerably smallare than what is seen in blazars. Finally, we should expect that thermal emission will be unpolarized, as it reflects the random motions of atoms and ions within the various emission regions, where there is not a preferred magnetic field direction with respect to us when the emission is integrated over $4\pi$ steradians of the source.

In practice, blazars are more complex. In the first place, the nonthermal emission from the jet will dominate in most bands. Secondly, no disk or torus has a single temperature, and when a wide range of temperatures is seen, the resulting spectrum will approximate a power law, so long as additional temperature components can be added on [13,14]. Thus, a firm detection of thermal emission in a given blazar *must* include a demonstration of a fit of a nonthermal component *and* a 'bump' in the spectrum, in which the observed points are well fit by either a single-temperature blackbody or a sum of multiple blackbodies. Ideally, it should also include information on the variability of both components. These requirements are things we will return to in later sections.

An excellent example of how this can be done can be seen in the first solid detection of thermal emission in a highly beamed blazar [16], namely the famous object 3C 279. In that paper, the authors used an historical minimum in the nonthermal emission to detect a 'bump' in the UV spectrum, which could be traced to a thermal origin. As can be seen, in that paper there are adequate points to fit both the nonthermal and thermal continuum components, and it is also demonstrated that the thermal component varies differently than do other parts of the spectrum.

Unfortunately, the real world is never this neat. Rarely do we have enough data to make this convincing of a case for thermal emission. Either the object is in a somewhat higher flux state, or we will not have enough data to independently constrain both components spatially or spectrally. This makes constraining thermal emission in blazars difficult, and necessitates the approach taken here. In the next section I will use the approach outlined above to assess various claims for the detection of thermal emission in blazars, both in individual objects and also statistically on samples.

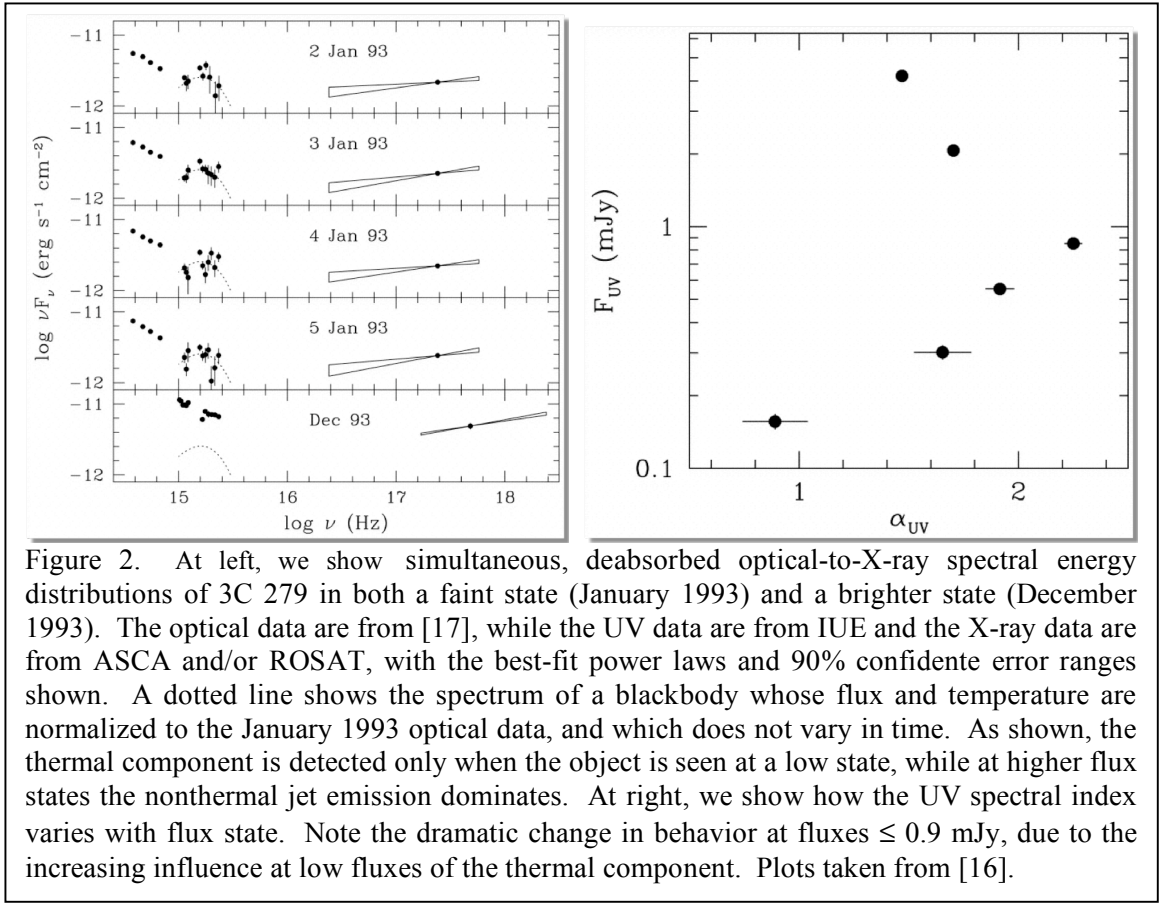

Figure 2. At left, we show simultaneous, deabsorbed optical-to-X-ray spectral energy distributions of 3C 279 in both a faint state (January 1993) and a brighter state (December 1993). The optical data are from [17], while the UV data are from IUE and the X-ray data are from ASCA and/or ROSAT, with the best-fit power laws and 90% confidente error ranges shown. A dotted line shows the spectrum of a blackbody whose flux and temperature are normalized to the January 1993 optical data, and which does not vary in time. As shown, the thermal component is detected only when the object is seen at a low state, while at higher flux states the nonthermal jet emission dominates. At right, we show how the UV spectral index varies with flux state. Note the dramatic change in behavior at fluxes ≤ 0.9 mJy, due to the increasing influence at low fluxes of the thermal component. Plots taken from [16].

### 3. What is the current state of the art?

When discussing thermal emission in AGN, one can discuss either the accretion disk or the torus. By far the most success has been found in attempting to discover emission from the disk, and so this is where I shall concentrate the discussion. In the spirit of the previous discussion, I have decided to split the current claims for thermal disk emission in blazars into four types: purely spectral claims, claims based on only variability information, combining the two and statistical claims. I will close this section by discussing attempts to detect thermal or thermal-related emission in other bands.

**3.1 Claims of Disk Emission Combining both Spectral and Variability Information**

We discuss first the case where both spectral and variability information can be combined. In this case it is relatively easy to demonstrate both of the requirements put forward in Section 2. We have already discussed one case where such evidence was presented. Here we discuss two other, more recent cases in the literature.

The first of these is shown in Figure 3 [18,19]. Those papers discussed the 2005 and 2006 multi-wavelength campaigns on 3C 454.3. What those authors demonstrated was that the UV emission of the object was, when in a faint state, of a very different spectral morphology than that seen in any other band. That component was not seen in higher synchrotron/Compton flux states. Closer examination of the SED in fact revealed a two-temperature shape, attributed to (respectively) the big and little blue bump components. Curvature was also seen in the soft X-ray, when the object was faint.

The second case I want to mention is the work of [20], which analyzed *XMM-Newton* observations of AO 0235+164. As in the case of 3C 454.3, time-variable spectral curvature was found in the object's UV-X-ray spectrum, which was most pronounced in low flux states. Unfortunately, the particulars of the SED resulted in a much smaller amount of the thermal component being visible; however, as five different flux states were observed the claim seems secure.

### 3.2 Purely Spectral Claims of Disk Emission

In general, most claims for individual objects fall into either this category or that in the next subsection. By their very nature, when one has only one sort of information, one needs to be rather more careful to not overinterpret one's data.

An example of a case where there is good evidence to support a claim for thermal emission can be seen in Figure 4, which shows one of four high-redshift blazars analyzed by [21], that had been detected in the 15-150 keV band with *Swift*/BAT. In three of the four objects, there was no evidence for variability detected either in the XRT or BAT. As shown, the XRT data for this object constrain well a nonthermal component that is increasing strongly as one goes to higher frequencies, and yet cannot be the same emission mechanism as that which occurs in either the radio (historical data) or UV-optical (UVOT). In fact, one cannot combine a single synchrotron emission component with its mandatory Comptonization to produce all three of the visible 'humps' and the most consistent explanation is that the synchrotron emission is seen at radio-IR frequencies with the UV and optical bands being dominated by a thermal component, attributed to the accretion disk.

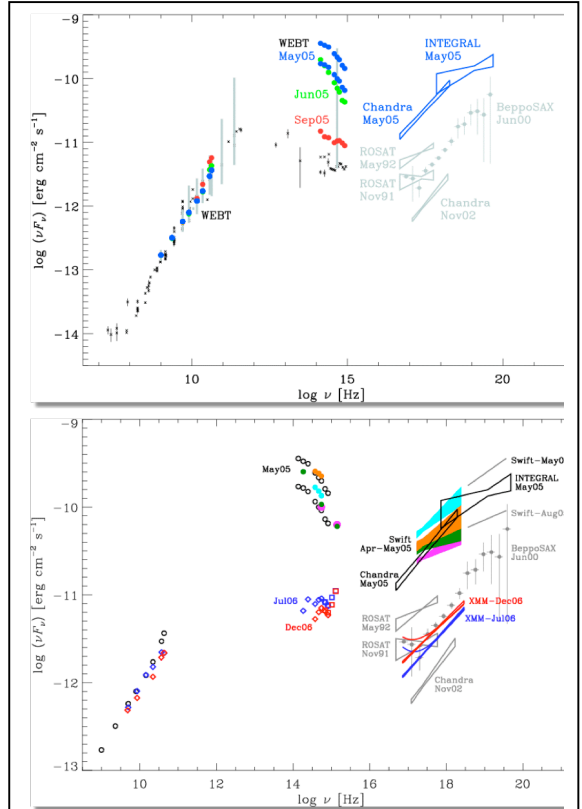

Figure 3. Spectral Energy Distributions of 3C 454.3 during several epochs in 2005-2006. As can be seen, in low flux states a thermal component was visible in the optical-UV part of the spectrum. Plots taken from [18] and [19].

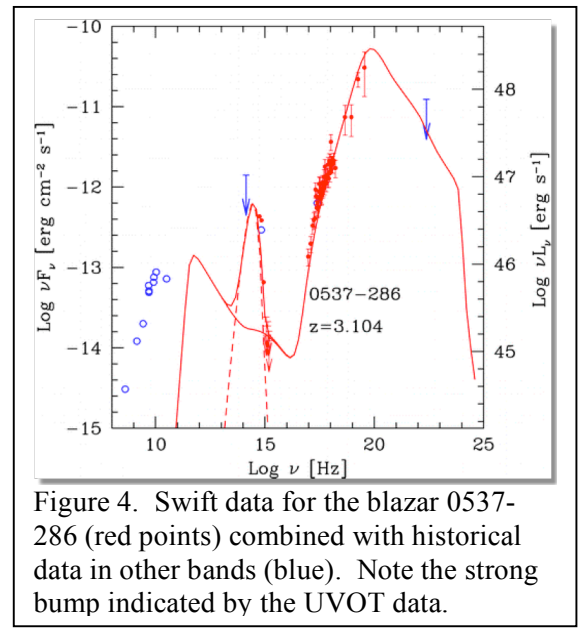

Figure 4. Swift data for the blazar 0537-286 (red points) combined with historical data in other bands (blue). Note the strong bump indicated by the UVOT data.

Not all claims live up to this standard. An example is the work of Landt et al. [22], who analyzed *XMM-Newton* and *Chandra* data on HFSRQs from the DXRBS. Along with the speaker, those authors had previously [23,24,25,26] found that some HFSRQs were not highly beamed and only 2 showed *bona fide* X-ray synchrotron peaks. In [22], that information, combined with new X-ray spectra, was used to model the UV-optical emission from all HFSRQs as a brighter-than-average accretion disk and corona. This was justified for some modestly beamed or unbeamed (i.e., *R*<1) objects where one could presume weak synchrotron emission. However, there were other objects where it was less well justified, i.e., objects with higher *R* values and possibly also sparse, nonsimultaneous SEDs (which in the optical included spectra and sky survey data, but just a single point in radio) that suggested large-amplitude, frequency-dependent variability (see Section 2). For reasons of space, I do not show those SEDs here; the reader is referred to Figures 4 and 5 of [22] for supporting evidence.

### 3.3 Claims of Disk Emission Based Solely on Variability

It is more difficult to establish, purely on grounds of variability, that the likely mechanism for the observed emission in a given band is due to thermal radiation. The reason for this is similar to the above: because of the dominance of the nonthermal emission, in order to obtain firm evidence implicating a different mechanism one must demonstrate widely different variability behavior in adjacent bands. And even in the best of circumstances, where one can demonstrate that (e.g.) fast variability is seen in higher and lower frequency wavebands but that the variability is much slower and/or nonexistent in the band dominated by the putative thermal emission, the claim is by necessity much less certain. The reason is that there may not be sufficient spectral data to show a thermal shape to the continuum in that band. For example, in [27], *RXTE* and ground-based observations of PKS 2155-304 showed for ~10 days, optical variability that was completely uncorrelated (although not on vastly different timescales than) that seen in X-rays. Those authors suggested that the most likely possibility was thermal emission dominating in the optical, although other interpretations were possible. Unfortunately, no spectral fit was possible with so few points, so it is not certain that we were not seeing a completely different region of the jet (although good arguments were presented to discount that possibility). The reader is referred to Figure 7 of that paper for the variability information.

### 3.4 Statistical Claims of Disk Emission

A number of authors [28,29,30] have used the general characteristics of blazar spectra, plus the relatively narrow-band nature of thermal emission in AGN, to argue that a substantial fraction of the UV emission in blazars comes from thermal processes. They had to utilize other evidence to estimate accretion disk luminosity, usually emission line data, bolstered in the case of [30] by the use of VLBI data to determine *R* and the broadband SED. Hence their results are somewhat model dependent. They achieved modest success, showing that a significant fraction of the UV luminosity (typically 10-50%) in many blazars was due to thermal emission from the accretion disk. However, the total accretion disk luminosities estimated by those authors varied by as much as an order of magnitude, mostly because they were unable to estimate the upper and lower wavelength ends of the band where the accretion disk dominated, and also because very few blazars have good estimates of black hole mass. This subject is promising, however, because the general result that ~10-50% of a typical blazar's UV emission comes from the accretion disk and/or corona, is broadly consistent with the results (Sections 3.1-3.3) from multiwaveband campaigns and other sources [16,17,18,19,20,21].

**3.5 Claims of Thermal Emission in other bands**

In other bands, one can look either for the thermal emission from the infrared, or line features that are generated directly within the disk. Regarding the former, there are is very limited evidence, and in fact ISO observations of the 1 Jy BL Lacs were generally well fit by a single synchrotron emission component [31], although a few objects showed slight evidence of IR excesses that could, however, be explained by variability.

As with searching for thermal disk emission, a more promising method is to combine spectral and variability information on a given object to search for this component. This has recently been done for the blazar PKS 1510-089 by Kataoka et al. [32]. In that work, which analyzes data from a 2006 August multiwaveband campaign, Kataoka et al. took X-ray and optical data from a variety of observatories, although the mid-IR data was historical data from IRAS. The broadband SED of that object, however, clearly showed evidence for two "bumps" above a synchrotron model, likely due to thermal emission: one in the optical-UV, as discussed in the previous sections for 3C 454.3 and other objects, but also a second in the IR which would likely come from the torus. This result is not as iron-clad as some of the others, unfortunately, given the fact that the IR data was historical and not simultaneous; however, it is a notable result and we should not expect strong variability from the torus emission as explained above.

There have also been recent observations in other bands that show promise. Sambruna et al. [33] detected Fe K$\alpha$ emission in *Chandra* observations of two blazars, PKS 1136-135 and 1150+497. This emission was moderately broad, consistent with the "X-ray Baldwin Effect" [34,35] that has been noted in other AGN classes. This line is believed to emerge from material orbiting in the innermost stable orbits of the accretion disk and is therefore a strong link to the presence of a thermally emitting accretion disk in these objects, although because the line is linked to the corona emission its luminosity cannot be used to estimate a disk temperature and/or luminosity. It should be noted that at least one of these two objects (PKS 1136-135) exhibits relatively modest beaming and possibly also evidence for X-ray synchrotron emission by its kpc scale jet [36], a characteristic also recently claimed for the jet of 3C 273 [37,38,39].

**4. Links to Other Parts of the Spectrum**

The evidence from the previous section can be summed up as follows: while the evidence is sparse, one can definitely make a strong case that there is thermal emission present in blazars, particularly emission line objects. The presence of thermal emission, particularly from the accretion disk, is fortunate, as most models to generate relativistic jets rely on the presence of an accretion disk [40]. Moreover, as discussed in Section 1, the presence of strong, broad emission lines in FSRQs is by itself evidence that the overall energy budget of the nuclear engine is dominated by the accretion disk rather than the jet. With the presence of such a strong, isotropic radiation field that can react with jet particles via Comptonization, it is important to discuss the effect of thermal – and thermally excited – emission on the observed spectrum.

We discuss these issues with the aid of the 'Compton Shoppe' models [41], which we have developed to be a publicly available toolbox for modeling synchrotron-inverse-Compton spectra. This toolbox, which is accessible on the Web at http://www.jca.umbc.edu/~markos/cs, will be discussed at length in a series of future papers [42,43,44]. It includes both single-zone and multi-zone, 'pipe' configuration models. Its main advantages are public availability, which includes extensive documentation, and also a fully self-consistent treatment of all cross-sections and physical mechanisms in a relatively user-friendly manner. Unlike many previous models, we fully include effects such as multiple order scatters, relativistic cross-sections and pair production. The results we discuss in this paper were run only in the single-zone, 'Compton Sphere' suite, which is currently the only one that is publicly available (the multi-zone 'Compton Pipe' suite should be available in fall 2008).

While we do not need to discuss the accretion disk itself as a source of external Compton seed photons (because those photons would come from the reverse direction in the source frame and so would be relativistically de-boosted), we do need to discuss the external Compton emission from other regions whose main power source is the accretion disk. The most obvious if these is the BLR, for which recent work has concentrated on the 'mirror models' for the generation of GeV gamma-ray emission in blazars, particularly emission-line objects such as 3C 279. The basis for these models was the analysis in [45] showing that the gamma-ray variability found in the 1993 and 1996 flares of 3C 279 was greater than quadratic – a result that has since been duplicated in other objects. Those authors claimed that such a result could not be duplicated by simple synchrotron self-Compton (SSC) models. Since that time, a variety of authors have discussed models that produce GeV emission by Comptonization of broad-line region photons (e.g., [46,47,48]). Those authors showed that by augmenting the seed photon field using BLR photons, one could reproduce the general outlines of the variability observed in 3C 279 and other blazars. Importantly, however, there have not been concrete attempts to model and compare with observational data both the variability and broadband SED of such a source in detail, most likely due to the paucity of detailed data in the gamma-rays. With this as backdrop, then, it is important to point out that it is *not necessary* to invoke a seed photon source external to the jet, to explain the GeV emission from blazars. The reason for this is that the statement that SSC cannot produce superquadratic variations is incorrect. The only regime where it is correct, in fact, is where only one scattering is considered which is correct only for the least luminous (and therefore least Compton dominated) gamma-ray blazars. In more luminous objects, the optical depth to Compton scattering increases (as it is proportional to the density of particles), and second and higher-order scatters become more important. These scattering reactions can, in more Compton-dominated objects, come to dominate the energy output from the SSC process. Importantly, when the high-order scattering reactions dominate, one *naturally* sees super-quadratic variability (Figure 5), similar to what was observed in 3C 279.

The physical inputs to that model include a bulk Lorentz factor of 5, a Compton dominance (ratio of synchrotron to Compton peak luminosity) of approximately 5 and a synchrotron peak frequency $\sim 10^{15}$ Hz before any perturbations are inserted. The first two parameters are typical of gamma-ray blazars, but the last is higher than observed in 3C 273; however, we have verified that the general result does not depend on this since the vast majority of the scatterers are still at low energies. And indeed, the general result of producing superquadratic variability from the SSC mechanism alone, can be produced both for objects having synchrotron peaks at IR frequencies (such as 3C 279) and those that peak at UV-X-ray energies (such as the TeV BL Lacs). An early version of these results was presented in [40].

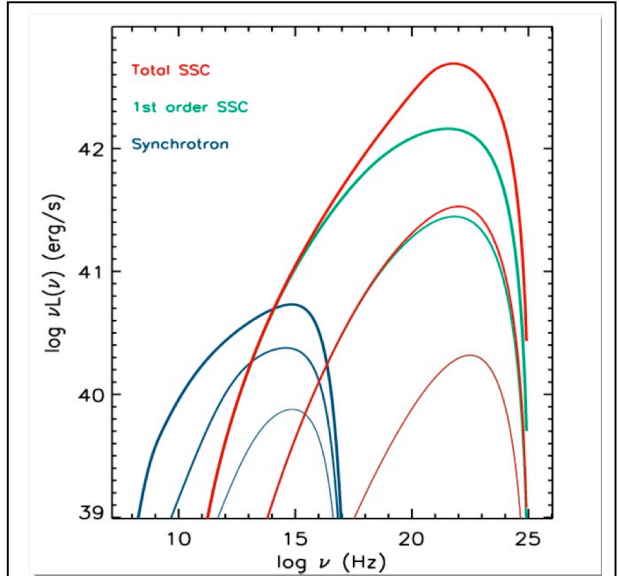

Figure 5. A model Compton-dominated source. In this source, no external photon field is assumed, and the intrinsic ratio of total synchrotron to Compton power is ~5, prior to perturbation (lower set of curves). The upper 2 sets of synchrotron curves represents variations in the synchrotron power by 3.2 and 7.1, while the Compton curves shown represent the response. The emitted power from both first and higher order scatters are shown. As can be seen, the superquadratic behavior is due to the contribution of the higher-order scatters.

Thus both SSC and EC models for the production of gamma-ray emission from blazars are currently viable, and it will require GLAST observations to decide between them. How can this be done? In Figure 6, we show a model run that was created with parameters that match those observed in 3C 279. A number of features can be seen in the resulting spectrum, including a 'hump' at optical-UV energies, produced due to the fact that for this choice of parameters, $\varepsilon_0 \gamma \sim 1$ for seed photons $\sim 10^{14}$ Hz. Thus, the scatters producing GeV photons occur in the Klein-Nishina regime. This has a few other important repercussions, namely that the GeV variability would be achromatic (as electron cooling is not energy dependent in the Klein-Nishina regime), and the GeV spectrum would be flat. Previous observations have not been able to test these predictions, since with CGRO data the measurement of slope in the gamma-ray spectrum was crude at best and it was not possible to measure variability timescales at multiple gamma-ray energies. What data does exist on the spectra of GeV blazars, however, shows that flat GeV spectra (photon index ~2) are not uncommon [49] among both BL Lacs and FSRQs. Achromatic variability *has* been measured in the synchrotron emission of some blazars, e.g., [50]; however it is uncommon. Importantly, the EC model depends on the assumption that the varying synchrotron source is located within the BLR *and* that the BLR photon field is relatively isotropic around the source. If either of these assumptions is incorrect, which can be the case if the BLR geometry is not spherical, as has been indicated by some groups, (e.g., [51,52]) the EC model becomes much more difficult to support. Along these lines the recent work of [53] is particularly notable. While they do not attempt to model a non-spherical BLR geometry, they have done the needed hard work of incorporating photoionization codes (e.g., CLOUDY) into their EC model. The overall SED shape from their work also approximates that shown in Figure 6, which does not include individual lines.

While there is not yet conclusive evidence for torus emission from blazars, it does need to be present at some level if unified models hold (particularly in the most powerful sources). It is therefore worthwhile to consider the impact of a thermal infrared photon field on the jet and the resulting inverse-Compton spectrum. A model spectrum is shown in Figure 7. As can be seen by comparison to Figure 6, this situation has a number of features in common with EC-BLR models, in particular the IR excess, which is partly a result of the onset of the Klein-Nishina regime, and also the "bump" in the

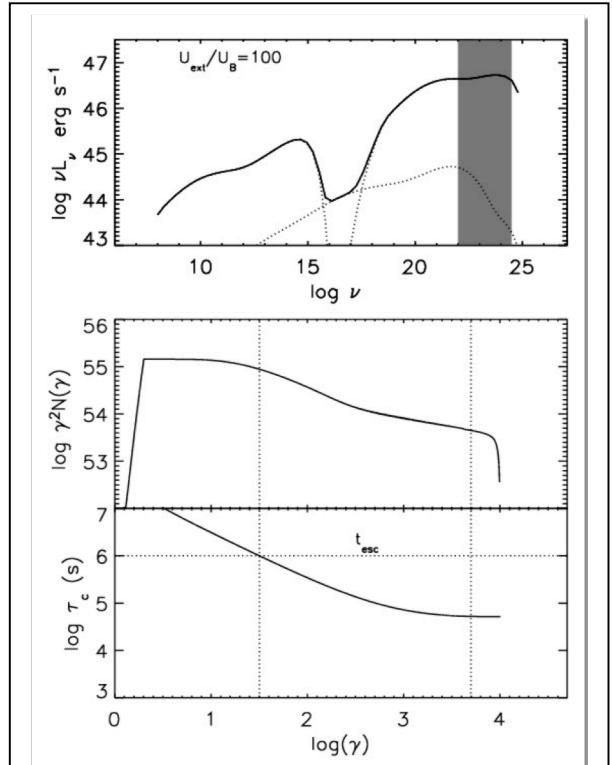

Figure 6. A model for an external Compton dominated blazar. Here the assumed ratio $U_{ext}/U_B = 100$ and we have assumed a source size $\sim 5 \times 10^{16}$ cm to ensure the dominance of EC over SSC. Bottom panel: the electron cooling time as a function of $\gamma$. Middle panel: the electron energy distibution. Top power: the emitted power. Solid line represents total power, and dotted lines represent synchrotron (leftmost), SSC (central) and EC (rightmost and most powerful) components. The gray band is roughly the EGRET-GLAST regime.

Compton component combined with the overall flat shape in the gamma-rays. This would be reminiscent of the gamma-ray spectra of the so-called MeV blazars (see [54] and references therein), particularly in objects that are more Compton dominated than the example shown in Figure 6. It should be noted here that the overall geometry of the torus would need to be roughly spherical (although perhaps patchy) for this model to be pertinent; however, there are a variety of current models that permit such a morphology for the torus, in particular the CLUMPY models [55,56]. And recently, de Rosa et al. [57] used the X-ray spectrum and SED of the blazar PKS 1830-211 to argue for Comptonized dust emission in that source, presumably with the seed photons coming from the torus [I have not included this above as it is not an example of a (claim of) direct detection of a thermal emission component, but rather a claim of detection of its Comptonized byproduct only.]

## 5. Discussion

We have shown that there exists now a considerable volume of evidence for the existence of thermal emission in blazars, at least in the case of the accretion disk. However, that emission is clearly difficult to detect, particularly if the object is not in a low state. This is in line with unified schemes, which predict that this emission must be present in all blazars, albeit often overshadowed (due to beaming) by the nonthermal emission. We have also shown that the existence of thermal emission in a given blazar – in any band – has unavoidable effects on the spectrum at higher energies, via Comptonization. It is important to attempt to isolate and gain information about the thermal emission in as many blazars as possible, particularly given the broad impact of this component both on the high-frequency

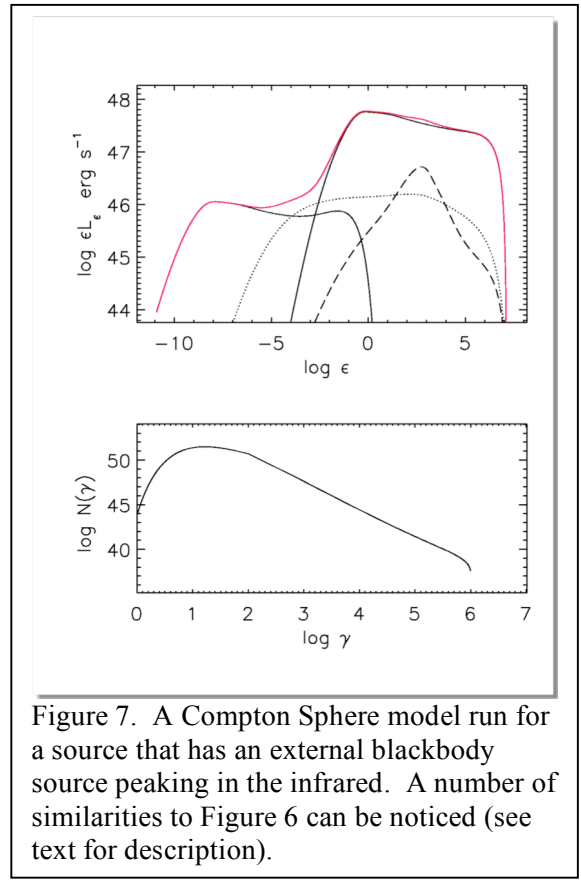

Figure 7. A Compton Sphere model run for a source that has an external blackbody source peaking in the infrared. A number of similarities to Figure 6 can be noticed (see text for description).

SED and variability as well as our overall picture of these objects under unified schemes. This also has a possible link to 'blazar sequence' scenaria, particularly given the likely importance of the disk emission in some HFSRQs [22,58]. Unfortunately we clearly do not have the data yet to clearly link this component to the blazar sequence.

Multiwaveband campaigns with broad coverage of many wavebands clearly represent the scenario under which one has the best chance to constrain both the thermal and nonthermal components in a given blazar. However, the situation is not at all hopeless in other cases. The examples given in Section 3 illustrate how important it is to constrain both of these components: not only is it improper to apply a 'one-size-fits-all' approach, but also without adequate information on the nonthermal emission that we *know* must dominate in most bands, Occam's razor dictates that it is improper to try to add a thermal component. We must also be careful not to apply any prior prejudices to what we do – i.e., the detection of superquadratic variability in the gamma-rays should not by itself be taken as evidence that the seed photons for high-energy Comptonized emission lie outside the jet.

Future satellite missions can help significantly in this regard, either by following the model of *XMM-Newton* and *Swift* in having UV-optical telescopes that are confocal with the main X-ray or gamma-ray detectors, or by expanding the range for multiwaveband campaigns into the infrared (*Herschel* and *JWST*). Future X-ray missions will also be able to look for Fe Kα emission (*Constellation-X, XEUS*). But also, GLAST will be able to help considerably in our efforts to pin down the thermal emission in blazars because of the unavoidable imprints they leave on the high-energy, Comptonized emission. As already noted, because the infrared to UV bands in many objects corresponds with the spectral region where the Klein-Nishina cross-section is what governs Comptonization, this produces unavoidable 'bumps' in the SED (Figures 6, 7). Thus with GLAST data, we will be able to deconvolve the thermal and nonthermal emission components *even in cases where the object is in a high state.* I should point out that if we really want to do the job right here, we also want to try to add optical/UV spectroscopy to multiwaveband campaigns, as the BLR photons that would be most likely to be Comptonized by the jet would be ones that lie within the jet's beaming cone – and thus one might possibly see line variability that could be correlated to jet variability events.

In closing, it is safe to say that the detection of thermal emission from blazars is a highly important subject, about which we are just beginning to learn. Future missions hold significant promise both in the direct detection of these emissions as well as their Comptonized byproducts. The importance of detecting this emission cannot be overstated. Not only can it help us constrain unified scheme models, where blazars are linked to unbeamed AGN where these thermal components are dominant, but also it can help us learn about the interplay between the disk and jet in AGN, a topic on which blazars have a unique view. In fact, given the critical role played by the accretion disk in the production of the jet [39], it may be that by constraining both the disk and jet emissions in blazars, and how they affect one another, we can learn about the overall process by which matter from the disk is funneled into the jet and how the jet is regulated by the disk and other thermally-emitting regions. This exciting prospect is worth future exploration in a number of different ways.